**Dissipation and detonation of shock waves in lipid monolayers**

*Shamit Shrivastava*

*Department of Engineering Science, University of Oxford*

*Draft 22$^{nd}$ February 2018*

**Abstract**

Lipid interfaces not only compartmentalize but also connect different reaction centers within a cell architecture. These interfaces have well defined specific heats and compressibilities, hence energy can propagate along them analogous to sound waves. Lipid monolayers prepared at the air-water interface of a Langmuir trough present an excellent model system to study such propagations. Here we propose that recent observations of two-dimensional shock waves observed in lipid monolayers also provide the evidence for the detonation of shock waves at such interfaces, i.e. chemical energy stored in the interface can be absorbed by a propagating shock front reinforcing it in the process. To this end, we apply the classical theory in shock waves and detonation in the context of a lipid interface and its thermodynamic state. Based on these insights it is claimed that the observed self-sustaining waves in lipid monolayers represent a detonation like phenomena that utilizes the latent heat of phase transition of the lipids. However, the general nature of these equations allows that other possible sources of chemical energy can contribute to the propagating shock wave in a similar manner. Consequently, the understanding is applied to the nerve pulse propagation that is believed to represent a similar phenomenon, to obtain a qualitative understanding of the pressure and temperature dependence of amplitude and threshold for action potentials. While we mainly discuss the case of a stable detonation, the problem of initiation of detonation at interfaces and corresponding heat exchange is briefly discussed, which also suggests a role for thunder like phenomena in pulse initiation.

**Introduction**

Lipid bilayers and membranes are conventionally considered as the matrix that host functional proteins and enzymes, and play the role of compartmentalizing the cellular architecture [1]. However, recent research has proposed a more active role for these membranes in cellular functioning, as waveguides for acoustic propagation, suggesting a possible role in inter and intra cellular communication [2–5] (Fig 1). Not only these studies have confirmed the existence of such pulses both theoretically [6,7] as well as experimentally, but have also exposed new physical phenomenon that can occur at interfaces. This article establishes the observation of detonation waves, i.e. acoustic waves that can utilize as well as liberate chemical energy available at the interface.

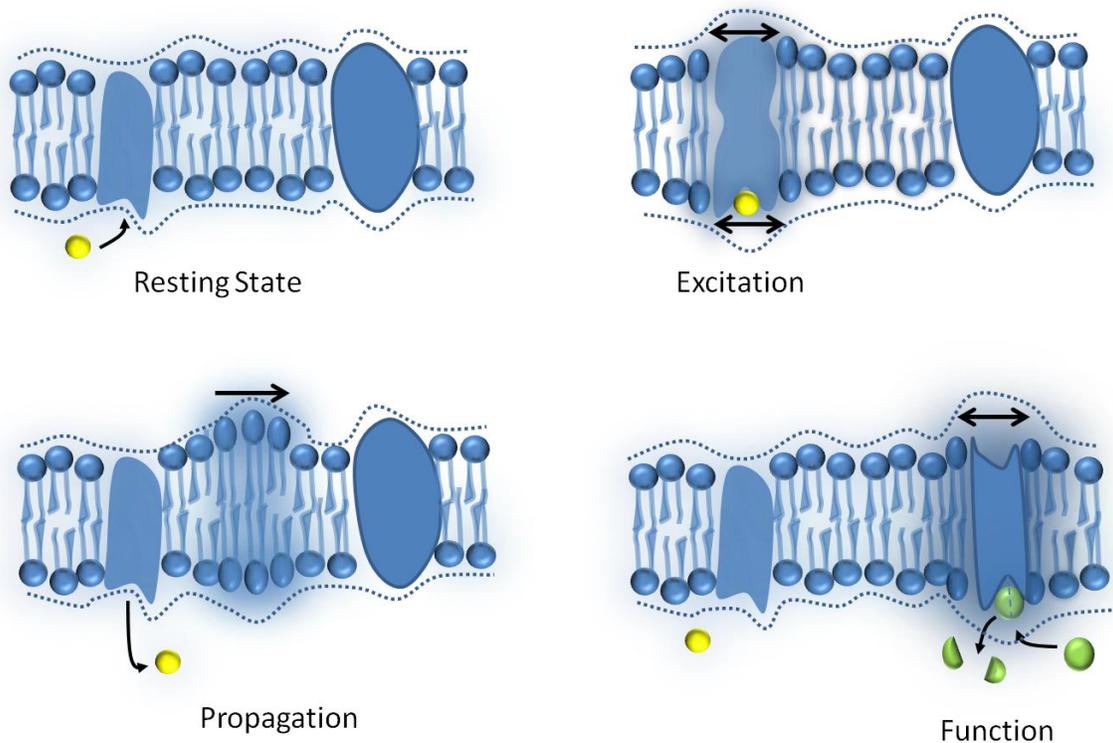

**Fig 1 Quintessentials of a signaling mechanism and how it might be felicitated via sound.** Conformational and heat changes associated with a ligand binding can lead to density perturbations at the interface which will propagate. The propagating density pulse might interact with a second enzyme altering its function and thus establishing a signaling pathway.

First and foremost, it has now been established that pulses at the interface originate from the same physical principles that result in propagation of sound, i.e. conservation of mass, momentum, entropy and constitutive relations [2,4–7]. However, this doesn't imply that they are simply density pulses, as sound waves are typically imagined to be. When the conservation principles are solved together with an incomplete constitutive relation, i.e. $P(V)$, it results in a purely mechanical description of the propagative phenomenon[7]. From a physical perspective, the constitutive relation is a description of the equilibrium state of the material, i.e. the state of maximum entropy $S(x)$ with $dS \sim 0$ and $d^2 S > 0$, where $x \equiv x_i$ represents all the observables in the system. Thus, when a medium is excited, by whatever means, the second law requires that the energy is distributed to all available degrees of freedom, which implies changes not only in volume or density, but also charge, polarization, ionization, magnetization etc. Hence, a sound wave propagating in a medium with all such degrees of freedom will manifest as a propagating perturbation in all the observables $x_i$ [4,8,9]. However, in order to solve the corresponding hydrodynamic problem for wave propagation (conservation mass, momentum and entropy), a constitutive relation of the form $S(E,V)$ or $E(S,V)$ is sufficient [10]. This relation can either be found on the basis of ab-initio calculation based on a molecular mechanical model or measured experimentally (e.g. steam tables). The latter constitutes the phenomenological approach and we believe is preferable, especially in complex materials such as biological interfaces, as it does not require us to make any assumption about the molecular nature and composition of the material. Rather the material is completely defined to a good approximation by the local specific heat ($C$) and compressibility ($k$).

The above discussion is true for any system and recently we have successfully applied these ideas to lipid interfaces based on a simple analogy between the entropy of 3D and 2D system[5–7,11]. In his 1901 publication on the phenomena of capillarity [12], A. Einstein began his article with the statement: "*If we denote by γ the amount of mechanical work that we have to supply to a liquid in order to increase the free surface by one unit, then γ is not the total energy increase of the system*." By applying a cyclic process to a water body with free surface, he deduced that an interface has its own specific heat and entropy that is independent from the bulk. Later, he laid down the foundation of statistical physics based on a phenomenological description of the entropy and the thermodynamic state of a system [13–18], which was fundamental for his work on Brownian motion, radiation, specific heats of solid and critical opalescence. Subsequently, in the 1980s, Konrad Kauffman realized that Einstein's phenomenological description of entropy, together with the fact that interfaces have their own entropy, also has implications for biological systems where interfaces are ubiquitous. Since entropy of the interface, when expressed as a function of area, also implies compressibility; phenomenon analogous to sound waves that conserve the entropy of the interface must also exists [19]. This insight formed one of the key elements for Kauffman's theory of nerve pulse propagation. In his work, the entropy potential of the interface forms the basis of both propagation as well as fluctuations (ion channels), which are claimed to be a consequence of the first $dS$ [20] and second derivative $d^2S$ [21,22] of the entropy respectively. Experimental validation of these predictions were carried out initially by himself [23,24] and later by others [2,5,9,25–28]. Particularly, it was shown that velocity of sound in a lipid monolayer at the air-water interface is determined by the two-dimensional compressibility of the interface. Also, as expected from the constitutive equations measured in terms of additional observables (fluorescence, pH, surface potential), corresponding changes in all these observables were also observed to co-propagate with the pressure wave. *Here, the interface is defined as the quasi-2D medium where the pulse propagates and it includes the hydration layer that* *envelopes the lipid monolayer.* We believe that lipids essentially allow us to control the state of this interface, which can be measured experimentally (surface pressure – area, surface potential – area, surface pressure – pH diagram), for example in a Langmuir trough[29].

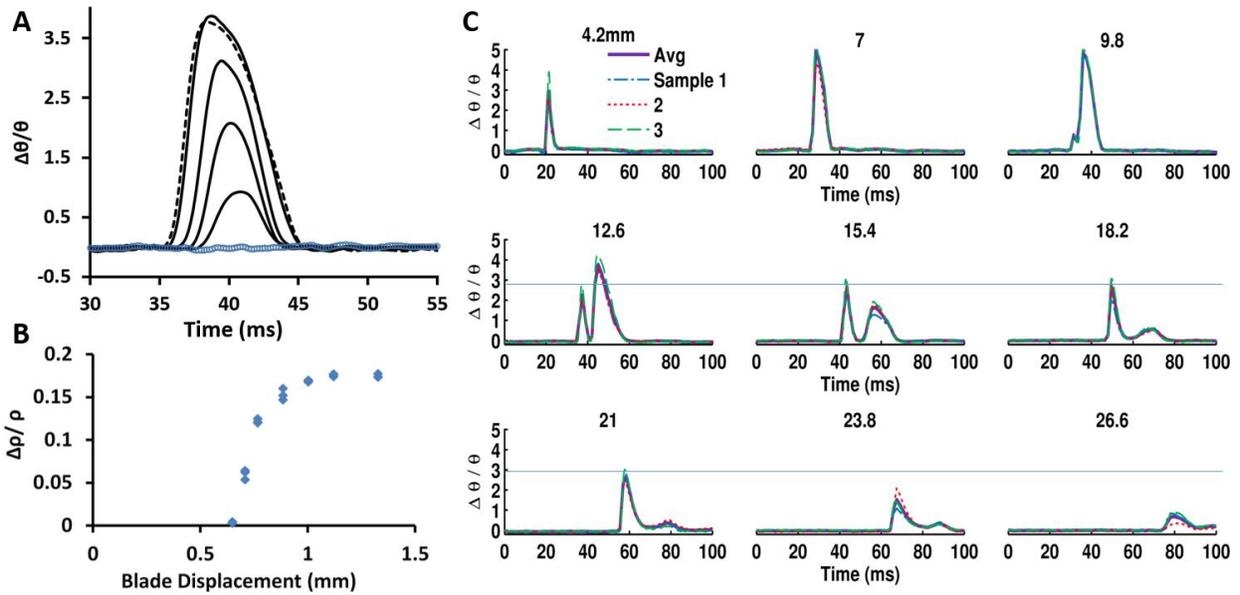

**Figure 2.** The key features of nonlinear pulse propagation in a DPPC (1,2-dipalmitoyl-sn-glycero-3-phosphocholine) monolayer at the air/water interface, as reported previously (Reproduced from ref. [11]; Copyright (2015) American Physical Society). (A) Pulse shapes as a function of amplitude indicates both the dispersion and nonlinearity as a function of excitation strength. (B) The amplitude response as a function of excitation strength shown the threshold and saturation effect. (C) The evolution of pulse shape as a function of distance show the effect of discontinuity in sound velocity at phase transition that splits the pulse into two. The figure also shows the self-sustained amplitude of for-runner pulse starting at 12.6mm till 21mm compensated by the energy from the slower condensation wave, indicating a detonation like phenomenon. The interface between the two pulses is in the metastable state as discussed in text.

Recently, it was shown (fig.2) that near a discontinuity or a nonlinearity in the state diagram, solitary nonlinear pulses can be excited in a lipid interface (fig 2A), that have a threshold for excitation and an upper bound on maximum amplitude (all – or – none) (fig 2B), are self-sustaining, and can annihilate upon collision [5,11,30]. Note that the discontinuity in the state diagram also implies a discontinuity in the compressibility and hence the speed of sound. Fig 2C shows evolution of one such pulse as a function of distance from the origin. As can be seen, the pulse splits into two at 9.8 mm, and from 12.6 to 21mm the forerunner one propagates with a constant amplitude while slower wave decays rapidly. The splitting of the wave is indeed explained by the discontinuity in velocity. Although these observations can be explained by nonlinearities and discontinuities in the *(surface pressure – area) diagram,* the constant amplitude of the forerunner wave could not be explained at that time, which seemed to propagate at the expense of the slower wave. Additionally, we have also shown that these nonlinear pulses can annihilate upon collision[30]. In both these cases the transfer of energy (heat) from slower wave to the forerunner during splitting or from the colliding waves to the environment needs to be considered. In other words, changes in the energy or enthalpy of phase change within the pulse need to be considered. In fact, understanding the role of heat is important for not only gaining a deeper insight into the propagation mechanisms, but for also addressing the larger questions regarding the excitation of these pulses by various means. Here we present a phenomenological thermodynamics based qualitative description of these observations [6,7] that suggests detonation like mechanisms are involved in the propagation of solitary shock waves at lipid interfaces near phase transitions.

Given the ultimate goal of applying these principles to biological processes, this work relates classical work in 1D theory of shock and detonation waves in terms of the corresponding state variables relevant for the membrane interfaces. Thus similar to the objectives of pioneering works in the fields of shocks and detonation in 40s [31,32], the objective here is to give a consequent and a qualitative theory that outlines the properties of the state diagrams that are necessary to maintain the propagation of shocks and detonation in the membrane. While we do not investigate the problem of the initiation of the wave-front (mechanisms of excitation) itself, the framework presented also addresses this to a certain extent and will be discussed briefly.

**Sound propagation in a lipid monolayer**

Let phenomenological potential $S(x)$ be defined as a function of extensive observables $(x \equiv x_i)$ such as volume, mass, charge, concentrations, energy in different compartments etc. Defined as such, $S(x)$ completely defines the macroscopic as well as the microscopic state of the system based on extensive observables via $w(x) = e^{S(x)/k}$, i.e. $S(x)$ is the entropy potential. Equilibrium is the most likely configuration of the system or a maximum of the entropy potential. This is true for each $x_i$, i.e. at equilibrium the entropy is observably maximum irrespective of the choice of observable/s. Any perturbation of the equilibrium $\Delta x_i$ results in a restorative entropic force $\sim \frac{dS}{dx_i}$, which represents the total derivative of $S(x)$ with respect to a generalized perturbation $dx_i$. Thus, unlike the usual assumptions of perturbing, say, only the volume, i.e. $p = -\frac{1}{T}\frac{\partial S}{\partial v}$, it is almost impossible to induce such changes realistically as they would not necessarily constitute the least path, which is properly captured by the total derivative.

For example, in an experiment the total restorative force that results from a change is volume is given by $\frac{1}{T}\frac{dS}{dv} = \frac{1}{T}\left(\frac{\partial S}{\partial v} + \frac{\partial S}{\partial q}\frac{\partial q}{\partial v} + \cdots\right)$. That is, the pressure measured by the instrument includes contribution from forces resulting from changes in, for example, charge $q$ as a consequence of a change in volume $v$ as well. This is true for the surface pressure measured by Wilhelmy plate in a Langmuir trough as well, i.e., $\pi = -\frac{1}{T}\frac{dS}{da} = \frac{1}{T}\left(\frac{\partial S}{\partial a} + \frac{\partial S}{\partial q}\frac{\partial q}{\partial a} + \frac{\partial S}{\partial \delta}\frac{\partial \delta}{\partial a} + \cdots\right)$ where $\delta$, the width of the interface, has also been included to emphasize that the measured surface pressure also includes contribution from the width of the interface and it hasn't been ignored. Elementary models, such as the ideal gas equation, calculate the total force based on the right hand side of the equation. Pressure derived based on an incomplete equation of state or a molecular model might ignore some contributions on the right hand side of the total derivative and hence such a model never provides the complete picture (violates the second law). Experimentally ascertained relations ($p - v$ or $\pi - a$ state diagrams) on the other hand directly measure the total derivatives. The perturbation and the resulting restorative force can either oscillate or relax [33] depending on the rate of heat transfer. Together with continuity equation and conservation of momentum, the entropy potential forms the basis for the propagation phenomenon. In the following, we revert to the classical theory of sound propagation where all the variables represent phenomenological quantities that should be measured experimentally, hence in the following discussion $p = -\frac{1}{T}\frac{dS}{dV}$ and its 2D equivalent being $\pi = -\frac{1}{T}\frac{dS}{da}$. Applying conservation of momentum on an element of the lipid monolayer we get; $\rho\left(\frac{\partial v}{\partial t} + v\frac{\partial v}{\partial x}\right) = -\frac{\partial \pi}{\partial x}$, where partials are with respect to time and space and not state variables[2].

**Classical theory of shock waves and shock waves in lipid monolayer**

The state of the fluid was the focus of some of the seminal works in shock and detonation physics where the qualitative aspects of the wave phenomenon (e.g. stability of shock front) were directly attributed to qualitative features of the EOS such as its curvature [32]. Our goal is to understand these principles and their role in dynamic processes at biological membranes, such as in the propagation of nerve impulses. In this regard, researchers have recently embarked on studying sound waves in lipid monolayer, a simple model for cell membrane that is ideal for a thermodynamic understanding of the wave phenomenon[2,6,9,34]. State diagrams of lipid monolayers at the air-water interface show remarkably rich behavior [29], making it a simple system where the principles of acoustics in 2D can be tested extensively. These experiments have not only led to the discovery of new acoustic phenomenon in lipid systems but have also provided crucial insights into the role of such considerations for a thermodynamic theory of nerve pulse propagation.

Let us start by considering the approximate scenario as elaborated in figure 3A. The wave front is assumed to already exist and moving with a constant velocity, $D$, in the negative x-direction in an interface between to mediums I and II. We assume the interface is adiabatically isolated from these medium during wave propagation (in general the coupling between the bulk and the interface decreases with an increase in frequency[7]). In other words, we ignore the heat conduction between the interface and the surrounding medium. The motion of the interface has a velocity u, due to compression, which is assumed to be only along the x-axis as well. Let $n_i$ represent the mole fraction of the chemical species $i$ that is associated with the interface. This includes everything from a change in the composition of the macromolecules and amount of various ligands or toxins bound to the interfacial receptors, to protons and ions that influence the interface and can change during a wave. However, when we only have nonlinear sound waves and not detonation, the material composition or $n_i$ remains constant by definition. From conservation of mass, momentum and entropy and using a constitutive relation in terms of enthalpy $h(p, v, n_i)$ (in line with ZND detonation equations), one obtains for a steady wave, a relation between pressure, $p$, and specific volume, $v$, [35] across the wave-front.

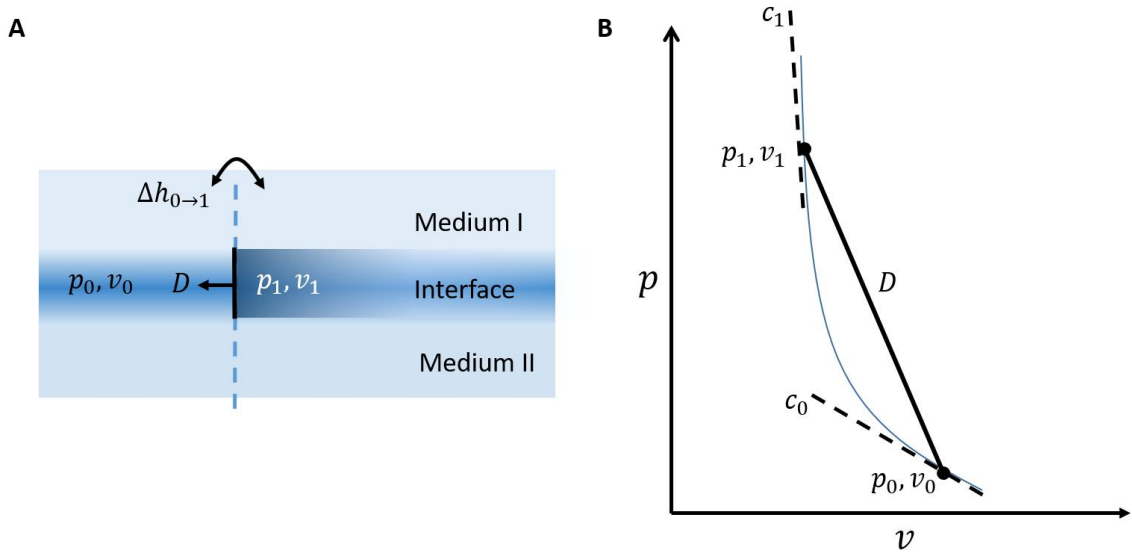

**Figure 3.** An approximate vision of an interfacial shockwave (the front). The interface is represented by the fuzzy band at the center, the effective width of which can be assumed to scale with the decay length of the wave away from the interface[7]. Irreversible heat exchanges take place in the reaction zone behind the shock wave, for example during a phase change, while the front represents an increase in enthalpy at constant composition (eq. 2).

$$D^2 = v_0^2(p - p_0)/(v_0 - v) = -v_0^2 \frac{\Delta p}{\Delta v} \qquad (1)$$

$$h - h_0 - \frac{1}{2}(v_0 + v)(p - p_0) = 0 \qquad (2)$$

For all practical purposes, in lipid monolayers, the surface pressure and area $(\pi - a)$ relation behaves as the 2D equivalent to the $p - v$ relation, even during extreme cases such as in shock waves at the interface [5,11]. However, in this article we will continue the discussion in terms of $p - v$ relations in order to maintain continuity with respect to the literature that this article builds upon.

Again, it is important to remind ourselves that all the variables here represent phenomenological quantities as discussed in previous section. The shock compression phenomena that we are trying to understand is the very process that can measure these quantities under the thermodynamic process of interest. Thus, unlike deriving shockwave characteristics from a theoretical equation of state, we are using the observations of shock waves in lipid monolayer to derive the characteristics of dynamic state diagrams of these systems and to understand the fundamental mechanisms. The observed shockwaves define the experimentally ascertained thermal behavior of the system (lipid monolayer) that help us interpret the nature of the corresponding adiabatic state-diagram and hence in future will allow us to arrive at the correct equation of state [7].

Now in eq.(1), for $\Delta v \to 0$ or $u \ll c$ it gives the relation between material properties and the velocity of sound $D = c_0 = (v/k_s)^{1/2}$ and $k_s = -\frac{1}{v}\left(\frac{\partial v}{\partial p}\right)_S$. Note that the partial derivative is only with respect to $S$, all other variables such as charge, ion concentrations, dipole orientations etc. that are coupled to volume changes are free to vary and contribute to $\frac{\partial v}{\partial p}$. In general, eq. (1) tells us that the initial and the final state across the wave-front should lie on a straight line in the $p - v$ plane where the slope is given by $-D^2/v_0^2$ (Fig 3B). It is very important to note that that the line corresponding to eq. (1) is not a thermodynamic path, it is only a mathematical tool and only its initial and final coordinates have real meaning, which comes from eq.(2). The initial and final points are given by the $p - v$ values satisfying thermodynamic criteria for stability lie on the state diagram, also known as the dynamic adiabat of Hugoniot. This is captured by eq. (2), which can be measured experimentally. As seen in Fig. 3B for $D > c_0$, eq. (1) makes a chord on the state diagram. If the state diagram is convex as assumed in the figure, then the velocity of sound increases continuously from initial value $c_0$ to final value $c_1$ across the wave-front or $c_1 > D > c_0$. *This results in a non-linear wave-front or a shock wave as higher amplitude part of the pulse will travel faster and reach the front ( $c_1 > D$ ) hence pulse gets steeper as it propagates.* This in turn increases the amplitude further at the front. In real systems, the accumulation of energy at the front is stabilized by dissipation and dispersion.

Acoustic waves measured in a lipid monolayer clearly show these nonlinear effects (fig 2A). The wavefront can be seen to arrive sooner as the amplitudes increase. Remarkably there also exists a threshold power beyond which the amplitude is significant and rises quickly before saturating at a maximum value (fig. 2B). Any further excitation and the pulse only broadens and does not increase in amplitude. In shock physics the broadening indicates an increased dissipation at the shock front as will be discussed below. Thus the saturation of the maximum amplitude already underlines the importance of energy or heat transfer. So why does the dissipation increase abruptly, which limits the nonlinear increase in the amplitude of the pulses? This requires an understanding of the role of phase changes that results in the nonlinearity in the state diagram. The observed nonlinear pulses represent a propagating phase change as discussed in the next section.

**Dissipation and phase change during a shockwave in a lipid monolayer**

The discontinuity or nonlinearity in the state diagram discussed in the previous sections results from the liquid expanded – liquid condensed phase change in the lipid monolayer. In particular, the velocity of sound *decreases* discontinuously as system goes from LE phase to LE-LC coexistence phase in a lipid monolayer. Hence, the convexity of the state diagram assumed in fig.3B is not true anymore which fundamentally affects wave propagation. Indeed, $\left(\frac{\partial^2 p}{\partial v^2}\right)_s > 0$ is known to be a necessary condition for stable shock fronts, derived in the seminal works of Bethe[32]. Intuitively, we can see why by considering the intersection of eq.(1) with a concave state diagram instead, thus we now consider $c_1 < D < c_0$. A sound wave of infinitesimal amplitude can exist along any part of the concave curve as before. However, for any significant amplitude, now the rarefaction or the negative pressure phase would propagate faster than the compression front. The pulse in this case self-interacts in a destructive way weakening the wavefront [35] and radiating it away as small amplitude sound waves. We have shown, both experimentally[5] and numerically[7], that the pulses excited in the concave region of the state diagram have negligible amplitude compared to the convex region. However, we have also shown that if the lipid monolayer is compressed faster than a threshold rate, large amplitude solitary shock waves begin to propagate whose velocity increases with amplitude [11], as discussed in the previous section. This indicates that the state diagram can behave as a convex curve above a certain compression rate.

So what happens at the threshold? Recall that the negative curvature or a negative discontinuity in the slope of the state diagram or eq. (2) results from the corresponding phase change. However, the rate of nucleation of a phase change has a finite value. Therefore, if the compression rate is faster than the rate of nucleation, then the phase change is not allowed kinetically. The system can then be compressed into a metastable or superheated regime without undergoing phase change. Because there is no phase change, the compression happens along a metastable state diagram (entropy is maximum with respect to volume but not internal energy), still satisfying $\frac{\partial^2 p}{\partial v^2} > 0$ resulting in a stable shock front. The rate of nucleation also increases as the system is pushed further into the metastable regime. The nucleation rate eventually diverges at what is theoretically knows at the spinodal boundary, where the phase change becomes unavoidable, thus destabilizing the wave front.

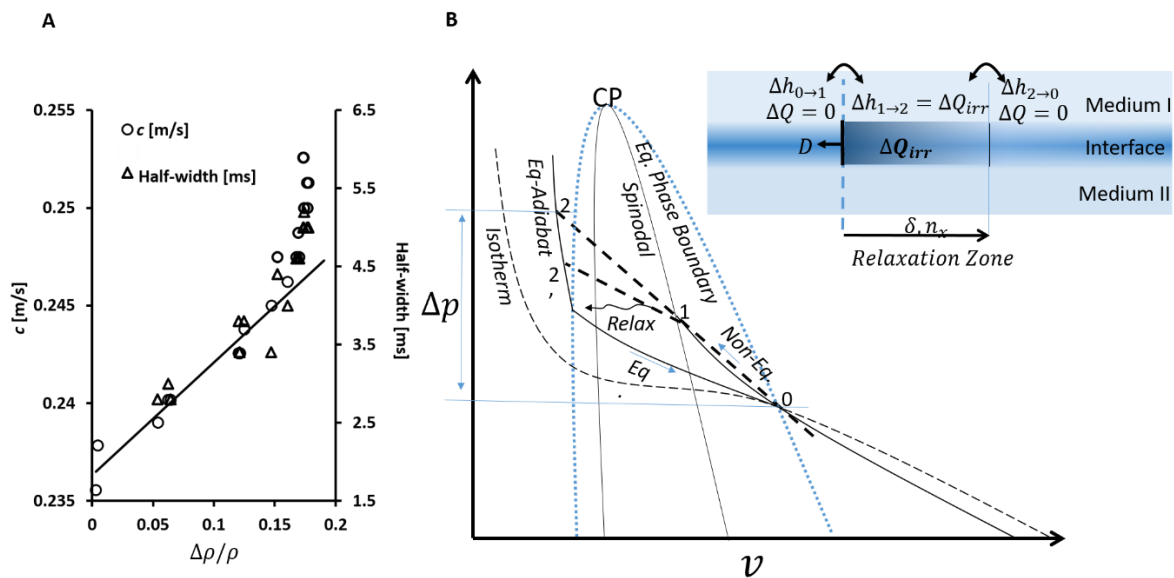

**Figure 4. Dissipation from phase change and relaxation** (A) Velocity and Half-width $\Delta t_{FWHM}$ of shock waves measured in a lipid monolayer as a function of relative compression. (Reproduced from ref. [11]; Copyright (2015) American Physical Society). (B) Corresponding states imagined on a $p - v$ state diagram. The pressure change of $\Delta p$ is the maximum pressure across the nonlinear wavefront. The equilibrium phase boundary or the saturation limit $\sigma$ marks the discontinuity in $\left(\frac{dp}{dv}\right)$ in LE vs coexistence region. If the compression rate is faster than LE→LC relaxation then the system proceeds continuously along non-equilibrium adiabat $0 \to 1$ without phase change. However, at the spinodal boundary phase change is instantaneous and system eventually relaxes $1 \to 2$ (or 2' when pulse splits) to the equilibrium adiabat $0 \to 2$. The inset presents an *approximation* for the corresponding state changes along a propagating impulse, where $n_x$ represents the fraction of lipids undergoing the LE-LC phase change. Here medium I and II are air and water respectively [7]. $\Delta Q_{irr}$ is the heat released during relaxation to state 2 that occurs over a distance $\delta \sim c_0 \tau$, i.e the shock or detonation width [36], where $\tau$ is the relaxation time.

The width of the shock-front provides crucial information regarding the nature of relaxation processes involved within a shockwave [36]. Figure 4(A) presents previously reported half-width of the nonlinear pulses measured in a lipid monolayer [11]. It can be seen that the width increases linearly with increased compression till $\Delta\rho/\rho \approx 0.15$. Then the width increases abruptly upon further attempts to increase the compression ratio. For weak shocks the width $\delta \sim c\Delta t_{FWHM}$ can be estimated as $\delta = 8aV^2/\Delta p \left(\frac{\partial^2 V}{\partial p^2}\right)_S$ [36]. Clearly, for $\Delta p > 0$ the relation is meaning full only when $\left(\frac{\partial^2 V}{\partial p^2}\right)_S > 0$. Here $a$ quantify dissipation in the system and is related to viscosities ($\eta, \zeta$) and thermal conductivity $\kappa$, $a = \frac{1}{2\rho c^3}\left[\frac{4}{3}(\eta + \zeta) + \kappa\left(\frac{1}{c_v} - \frac{1}{c_p}\right)\right]$, $c_v$ and $c_p$ are specific heats at constant volume and constant pressure respectively. Accordingly, for a constant $a$ the width should decrease with increasing amplitudes. However, we see an increase in the width that eventually diverges resulting in the saturation of amplitude. This indicates an abrupt increase in $a$ and/or a decrease in $\left(\frac{\partial^2 V}{\partial p^2}\right)_S$, which can be primarily attributed to an increase in $\zeta$ due to relaxation processes associated with the phase change [37]. For small amplitude adiabatic changes in pressure and density $\zeta = \tau\rho(c_\infty^2 - c_0^2)/(1 - \omega\tau)$, where $\tau$ is the relaxation time of the conformational change, $\rho$ is density, $c_\infty$ is the velocity of sound at the fixed non-equilibrium state, and $c_0$ is the velocity of sound at equilibrium[38]. Clearly as the timescale of the acoustic process becomes comparable to the timescale of the relaxation process, $\zeta$ diverges. Indeed, among the 5 different timescales reported in DPPC (1,2-dipalmitoyl-sn-glycero-3-phosphocholine) vesicles [39], two of them (time scales of phase separation and domain formation) correspond to 4-40ms range, and the observed $\Delta t_{FWHM} \approx 4ms$ is right in this range. Thus we can now draw a clear picture of the state changes during the observed shock waves, which will now be discussed.

As shown in figure 4B, beyond the spinodal boundary $\frac{\partial p}{\partial v} > 0$. Hence, the material becomes unstable and it relaxes $(1 \rightarrow 2)$ to the equilibrium adiabatic diagram, releasing the heat of transition in the process, which limits any further increase in amplitude and broadens the shock front as discussed. While the phase change is initiated at $(p_1, v_1)$ further compression travels at a slower velocity as $c_2$ in the phase coexistence region is $< c_1$ for the superheated fluid. Thus, dispersion $c(\omega)$ which is closely related to $\zeta$, also contributes to limiting the amplitude and the increase in the width of the impulse. In fact, the shock now consist of two wavefronts given by $D_{(0\rightarrow 1)}$ and $D_{(1\rightarrow 2)}$. The saturation amplitude is given by the straight line $D_{(0\rightarrow 1\rightarrow 2)}$ corresponding to the saturation amplitude. As the front propagates, dissipation reduces the compression rate and as a result $D_{(0\rightarrow 1\rightarrow 2)}$ splits into two wave-fronts given by $D_{(0\rightarrow 1)} >$ $D_{(1\to 2')}$. Note that even if the initial compression rate was less than the saturation limit but greater than threshold, then the phase change can still occur if compression rate goes below instantaneous nucleation rate, resulting in what is known as weak splitting [10]. The observed shock induced phase change in the lipid monolayer are exothermic. Hence $h - h_0 < 0$ in eq. (5), which means the energy released from phase change and relaxation can further compress the material, reinforcing the pressure increase in the shock front. The next section explains the mechanism which is essentially the phenomenon of detonation.

**Detonation in a lipid monolayer**

Changes in the internal energy of the medium during propagation, for example due to a phase change, are addressed by the detonation theory. In detonation theory, the accumulation of energy at the shock front due to $c_1 > D$ also forms the basis of a stable detonation wave, which can be explained as follows. The shock front represents an adiabatic compression during which both pressure and temperature rise. If there are reactive species in the propagation medium, the increased pressure and temperature of the final state $(p_1, T_1)$ can activate a chemical reaction. If the reaction is exothermic, the energy released will also propagate with the velocity of sound $c_1 > D$ and can compensate for any heat loss at the front, which propagates at the slower velocity of $D$. Since the chemical reaction alters the composition of the propagation medium, enthalpy in eq. (2) also becomes a function of composition apart from pressure and volume. Then, eq. (1) connects the initial and final states such that it intersects all Hugoniots that correspond to the intermediate states representing all intermediate chemical composition, $n_i$. As there is a family of curves that the chord D has to intersect, the unique value of D is based on certain criteria, which established that $D = c + u$. Details of the derivation can be found in the works of Zeldovich and Neumann [31,35]. The graphical method of Neumann (fig.5) considered two generic families of intermediate state diagrams and following two criteria were derived;

1. If the intermediate state diagrams do not intersect each other than the unique value of D corresponds to the line drawn tangential to final hugoniot. In this case, a tangent to the state diagram corresponding to final composition satisfies intermediate state diagrams as well. (Chapman – Joguet condition).
2. If the intermediate state diagrams intersect each other, then there exists an envelope for the family of curves, and a unique D is found by the tangent to the envelop.

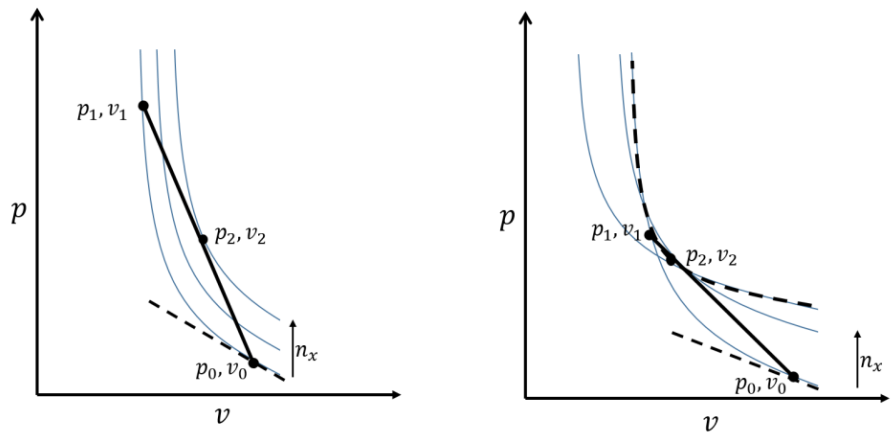

**Figure 5.** Two families of shock Hugoniots as discussed in the works of Neumann. On left the dotted curve represents the envelope for when the state diagrams intersect each other. The dashed straight lines mark the slope which is related to the sound velocity at the initial state. While D is related to the slope of the bold line as given by eq. (2)

Note that for a system where $n_i$ are fixed, D gives the shockwave velocity. However, as discussed, the lipid composition changes in the sense of a phase change during the shock. As the phase change introduces a discontinuity, it can also be imagined as two intersecting state diagrams. In the equilibrium case, the intersection lies at the phase saturation boundary (Figure 4B), which is the lowest pressure at which phase change can occur and coincides with the shoulder of the experimental isotherm. The point of intersection with the highest pressure is the point of termination of the non-equilibrium "dry" adiabat $(0 \rightarrow 1)$ on the spinodal boundary. All the intermediate state diagrams and their point of intersections lie between these two pressures. Therefore, according to condition 2 the maximum possible detonation velocity and amplitude is given by the line $D_{(0 \rightarrow 1 \rightarrow 2)}$ drawn tangent to the metastable state diagram with point 1 at the spinodal boundary. Thus, condition 2 of Neumann also provides an explanation of the saturation of the maximum amplitude from previous section observed for shock waves in lipid monolayers.

We also begin to see the connection between shock waves near a phase transition, as discussed in the previous section, and detonation waves. The shock front $(0 \rightarrow 1)$ takes the medium to a metastable state followed by a narrow zone of relaxation process $(1 \rightarrow 2)$. In case of detonation this is equivalent to the relaxation of chemical reaction to an equilibrium state [10]. The heat released at 1, propagates with velocity $c_1$ in either directions as sound waves or can potentially be emitted as photons [40]. The energy that propagates to the front reinforces the shock front (velocity $c_0 < c_1$), while the other part leaves the wave-packet as trailing acoustic radiation. This phenomenon is evident in the lipid monolayer as the propagating shock wave splits and the transfer of energy from slower wave to the fore-runner wave can be clearly seen (Fig. 2C). While the forerunner wave propagates with a constant amplitude the slower wave continuously decays. That the two waves propagate at different velocities also suggests that the compression (0 → 1) and release (2 → 0) represent different adiabats. Area enclosed by the process (0 → 1 → 2 → 0) underlines the dissipative nature of the process. Thus, heat released from phase change maintains the amplitude of the forerunner wave while being partially dissipated.

The detailed mechanism of detonation of shockwaves propagating in 3D mediums, by the latent heat of the phase transition, has been described previously by others [41]. Fig 2C shows that the phenomenon can indeed be observed in a lipid monolayer as well. Furthermore, the general nature of the discussion so far also suggests that exothermic reactions in general can couple to a propagating shock wave at a hydrated interface. Thus the nonlinearities in the experimentally measured state diagram that were shown to be the basis of the solitary pulse shape and all-or-none excitation; have now been shown to also form the basis for detonation. While in Fig 2 and Fig.4, the nonlinearities arise from the liquid expanded-liquid condensed phase transitions in the lipid monolayer, the characteristics of the propagating pulse are determined by the conservation of mass, momentum, and entropy alone. Therefore, in general the nonlinearity can be due to any other conformational transition, pKa or pK$[Ca^{2+}]$ etc., the reported observations are expected to be invariant of the origin of the nonlinearity and hence should be applicable in any case that presents itself, including a nerve.

Similar mechanisms likely form the basis of annihilation of colliding shock waves in lipid monolayers reported previously [30]. As the heat of transition contributes to the compression of the material and sustenance of the propagation, when such a pulse penetrates another, the heat of transition has already been released and is not available for the compression of the material ahead. As a result, the transmitted pulse will experience a significant drop in amplitude. These observations related to annihilation will be presented and discussed in more details elsewhere.

The above discussion also makes specific predictions that in principle can be tested experimentally. Relative to the state of a lipid monolayer, these shocks represent intense compression [40], especially when compressed beyond saturation and during collision. The mechanisms as discussed above suggest strong dissipation of energy under these circumstances which can possibly be detected as photo or acoustic emission, similar to the observations during the collapse of a bubble in cavitation [42]. Collisions of action potentials in large nerve bundles that support fast action potential (hence significant dissipation over a short duration) are most likely to present such evidence [43].

**State diagrams of excitable cells**

In previous sections, experimental observations in lipid monolayers have established a role for detonation and shock physics in pulse propagation at hydrated interfaces. Now we consider the extension of these results to excitable cells in general. A significant body of work on the excitability of living and nonliving systems provides the scientific rationale that allows this extension. As early as in 1902, Jagadis Chandur Bose presented his findings on the electro-mechanical response in the living and nonliving systems [44]. He observed that a wave of molecular disturbance under an electrical, mechanical or photo stimulus is accompanied by an electrical and/or mechanical response in animal nerves as well as in plants and metals. He further showed that these pulses change in a very similar manner even in such disparate system upon varying the environment, including the effects of temperature, chemical, anesthesia and toxins, suggest an underlying unification in physics. It was concluded that same physical laws that restore the equilibrium in inorganic systems are responsible for the phenomenon of response in living systems as well. Since then others, and most prominently Ichiji Tasaki, have repeated more sophisticated versions of these experiments [45]. From a material perspective, the only component common between these living and nonliving systems is the hydrated interface. Thus, it can be concluded that a general theory of response of hydrated interfaces is required to fundamentally understand the phenomenon of excitability in living system. This where the understanding derived from lipid monolayers becomes crucial.

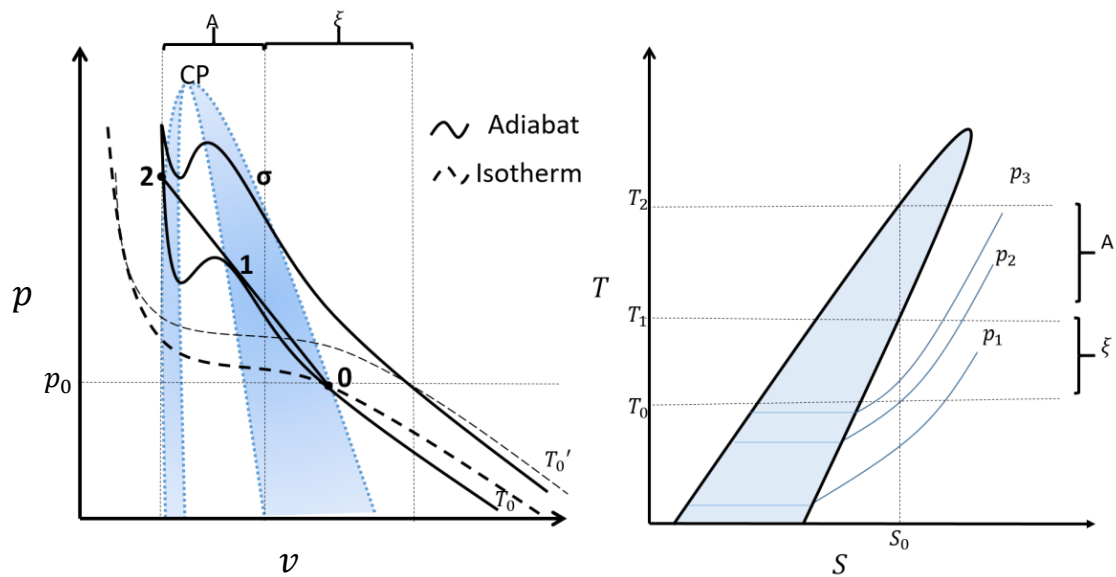

**Figure 6.** A PV and TS diagram as an aid to the eye, informed by experimental observations on lipid monolayer applied to an excitable membrane [46,47]. Coexistence region of the hypothesized transition region is shaded. Assuming that complete phase

change takes place across the shock front where the system is always in local equilibrium during an impulse starting at $(T_0 S_0)$, the wave front is represented by the isentrope $S = S_0$. Then the threshold and amplitude should scale with $\xi$ $and$ $A$ respectively, i.e. the distance between the initial state and the intersection of the isentrope with the phase boundary. $\sigma$ represents the saturation boundary.

However, as discussed above and as also stated explicitly by Neumann in his work on detonation that *the compressibility and the specific of the material under detonation are the decisive factors in the theory* [31]. Herein lie the challenges for a more quantitative application of eq (1). and (2). in excitable cells at present. This is because of our limited understanding or unavailability of the corresponding state diagrams, required in detonation theory. However, their importance for the nerve impulse phenomenon has been clearly emphasized in the extensive works of Ichiji Tasaki [48,49] and many others [50]. While membrane potential has been investigated extensively, the initial and final state of various observables across the nerve impulse wave-front, as a function of pressure and temperature at various concentrations of, for example, calcium, sodium and potassium need to be obtained systematically in an excitable system.

A theoretical framework, even if qualitative, is therefore necessary to guide and motivate such experiments. For example, a common misconception regarding the necessity of phase transitions in the thermodynamic approach is that a point-like phase transition temperature does not explain the marked temperature dependence in various characteristics of the nerve pulse phenomenon, such as threshold, amplitude and conduction velocity. This can be addressed in following manner. As discussed in a previous work[11], materials that consist of long chain hydrocarbons exhibit retrograde behavior (condense upon adiabatic heating) and have a typical TS diagram where it is simpler to represent these processes compared to corresponding PV diagram (fig. 6A vs fig. 6B respectively) [47]. The melting curves that are measured in a laboratory, usually represent an isobaric curve or one of the constant pressure curves. We get a precise line (or a point on $T$ axis) of constant $T = T_m$, intersecting the coexisting region at constant pressure (curve $p_1$). However, in the thermodynamic approach, the phase change during the nerve impulse can be approximated by a constant entropy process to first order. As one can see, this process cuts the coexistence region at entirely different points ($T_1$ $and$ $T_2$) compared to an isobaric one. Then as a first order approximation, for a pulse starting at $(T_0, S_0)$ the threshold for excitation should scale with $(T_1 - T_0)$ and amplitude of the impulse should scale with $(T_2 - T_1)$. Again same as $T_m$ measurements, the temperature studies on nerve impulses have been performed under isobaric conditions. Without the loss of generality, for the kind of TS diagram shown here, it can be seen that the threshold would increase while amplitude would decrease with increasing temperature at constant pressure; a typical behavior for nerve impulses [51]. On the other hand, pressure effects have been investigated under isothermal conditions. This is represented by moving the initial conditions from $p_2 \rightarrow p_3$ along the line of constant $T = T_0$. Again for the TS diagram shown here, increasing the pressure from $p_2 \rightarrow p_3$ will increase excitability as well as amplitude [52]. The conduction velocity (assuming weak shocks) will scale as $c(p,T) \sim (\rho k_s)^{-1/2}$ [46,53]. Thus a TS diagram of lipid like system can in principle explain general characteristics of pressure and temperature dependence of amplitude, threshold and conduction velocity in nerves. Figure 6 is for illustrative purposes only and we do not claim that it is a true representation of the state diagrams for a nerve membrane, but this is precisely why we need independent methods to measure such state diagrams in nerves in order to make these predictions quantitatively.

Motivated by the need to measure state in native environment, a relationship between the emission spectrum of membrane sensitive dyes and the specific heat of the interface was recently established [27]. Such dyes are widely used to measure the change in their spectrum properties during an action potential. Observed spectral changes [54] based on the recently established relation between the spectrum and specific heat predict stiffening of the nerve upon excitation, which is supported by independent observations based on X-ray diffraction studies [55]. Similarly, change in steady state fluorescence emission as a function temperature in excitable membranes show clear discontinuity in the state diagram near the so-called cold and heat block [56,57]. Furthermore, the presence of calcium in the media seems essential for the discontinuity corresponding to the heat block.

**Remarks on initiation of Impulse and energy changes during an action potential**

While the discussion so far assumed a steady state shock wave or detonation, a few remarks regarding the initiation of a shock impulse or detonation at a biological membrane can be made at this point. Initiation of an impulse, as demonstrated in lipid monolayers by an electrical [34,58], mechanical [5,46], chemical [4] or optical stimulation [59], are certainly possible mechanisms for excitation of impulses in cells as well. Each of these input stimulus that excite the interface are dynamic (i.e. they represent a dynamic external field that agitates the system). However, nerves can be excited by quasi-static changes in state as well, brought by a low constant current source or by exchanging ions [49,51,52]. To excite a shockwave, particle velocities on the order of the sound velocity in the material are required. The question remains: how a constant current source can abruptly kick the surrounding particles to the velocity of sound? To address this, one can draw inspiration from thunder shocks emitted from an electrical discharge in the atmosphere [60–62]. The discharge paths are referred to as "lightning channels", where they seem to be facilitated by dust particles, water droplets and other impurities in the environment. Once a conductive path is stablished by statistical coincidence, the atmosphere discharges locally. This results in immense heat released in a very short amount of time causing adiabatic compression of the air to high pressures in the vicinity of the channel. This pressure is then released as shock waves.

A 50mV potential difference across a 5nm lipid membrane creates a field strength of $10^6 V/m$, which is incidentally has the same order as the dielectric strength of various oils, which are usually likened to the hydrophobic core of the lipid billayer. The relation between adiabatic heating of the membrane due to ion flow has been discussed previously in a different context [63]. In addition, a quasi-static phase transition in lipid membranes have been shown to result in sudden increase in conductivity [64,65] and thus can cause sudden heating through abrupt ion flow across the membrane. Such processes can not only initiate a shock wave but subsequently channel the heat into the shockwave resulting in a stable detonation wave. From a detonation point of view, the enthalpy associated with any chemical change at the interface, be it due to mixing, absorption or cleavage of a substrate, can interact with the energy of the propagating wave and can be accounted for without the need for microscopic details of the chemical reaction. If we assume that enthalpy of mixing is insignificant compared to enthalpy of ion exchange, then energy entering the shockwave is represented by the corresponding ion concentration dependent family of state diagram $h(p, v, n_i)$ [4,66–68] as given by eq. 1 and 2.

While dielectric discharge at the membrane is one possible and very likely mechanism, the detonation can be initiated by any other form of a chemical process as well that can provide heat in short bursts. This could be relevant for the action of very fast enzymes, such as Acetylcholine esterase, which hydrolyses the neurotransmitter acetylcholine with a turnover rate of the order of $10^4$ with an enthalpy of the order of 8 kJ/mol [69], depending on the state of the system [70].

Finally, with regards to an understanding of how energy can be exchanged adiabatically between a lipid and its environment, recent experimental observations on the thermodynamic state of the interface during cavitation are very helpful, which will be published elsewhere [71]. When a suspension of lipid vesicles is subjected to shock expansion (in bulk water), cavitation or water to vapor transition ensues at the lipid interface. The state of the interface was measured simultaneously using optical probes, and it was shown that enthalpy flow associated with water to vapor transition across an adiabatically decoupled lipid interface is directly coupled to an equivalent enthalpy change within the interface, condensing the lipids in the process. In this sense, the experiments showed that heat released or absorbed at the interface can be channeled into the interface during an adiabatic process. In particular, an increase in entropy across the interface results in a decrease in entropy within the interface. This decrease in entropy can hypothetically compensate for dissipation from the propagating wave in accordance with the detonation theory as discussed.

At this point the observed heat changes during an action potential can be addressed. Quoting Abbott and Howarth [72]," *heat studies cannot be expected to themselves give rise to a model of the mechanism. The value of heat studies is that they represent a good criterion for assessing the fidelity of a proposed model system.*" Indeed, heat studies have consistently underlined the incomplete understanding of the phenomena of nerve pulse propagation. Measurements of heat exchanges during an action potential show a predominantly reversible character. The main predicament has been reabsorption of heat during relaxation which could not be explained by merely considering the capacitive currents (condenser theory) as pointed by several authors [72,73]. Note that in the detonation theory the reabsorption of heat occurs by default as the conformational changes relax back to initial equilibrium. Indeed, it was Margineanu and Schoffeniels [73] who showed that it is necessary to consider heat changes associated with reversible conformational changes in the ionophores to obtain a qualitative agreement between theoretical estimates and experiments. Still, the heat dissipation estimated from ionic currents remains significantly greater than the measured dissipation[73]. From the perspective of detonation theory only a part of the electro-chemical gradient across the membrane is dissipated as heat as rest feeds back into the shock front and hence is consistent with heat studies. Margineanu and Schoffeniels disregarded that the energy from ionic gradient can be utilized into conformational change in the absence of a possible molecular mechanism. Detonation theory provides this mechanism at least when considering the transitions in the nerve membrane from the macroscopic perspective[74,75]. As the pulses in Fig.2 also account for loss in amplitude due to spreading out, which is potentially also complicated by spatial focusing effects due to high nonlinearity, it is not possible to estimate the dissipation rate quantitatively based on these experiments alone. Observation of such pulses in confined geometries, on the other hand will likely allow such an estimate.

**Remarks on detonation vs Ionic hypothesis**

In detonation, external energy can be supplied to the propagating wave by chemical means. Therefore, it is important to make a clear distinction between the thermodynamic approach proposed here and the Hodgkin and Huxley model for nerve impulse [76]. In this regard, it is best to quote Neumann on the assumption he made before deriving the velocity of detonation [31]. *"We are disregarding the possibility that the detonation is propagated by special kinds of particles (ions, etc.) moving ahead of the wave head.*

*Indeed, if the views which we propose are found to be correct, no such special particles will be needed to explain the detonation wave"*.

In a similar spirit, Zeldovich, who arrived at detonation theory independently, emphasized [35]: *"Finally, entirely inadmissible at the present time are the attempts to identify the velocity of detonation with the velocity of motion of any particular molecules, atoms, or radicals in the products of combustion, the corresponding particles being assumed active centers of a chemical reaction chain. However good the numerical agreement, such an attempt is no more than a make-shift and a clear backward step with respect to the thermodynamic theory as is evident from the fact alone that it is entirely unclear what mean or mean square velocity, or other velocity of the molecules, should enter the computation."* Thus, in detonation, the shock wave precedes chemical changes and the propagation of the front is acoustic, unlike the Hodgkin and Huxley model where the front is carried forward by charge particles.

From a historical perspective, it is interesting to note that the detonation and shock theory were developed in parallel to the Hodgkin and Huxley model in 1940s [31,35]. However, the former was mostly classified defense research. While most of this research was eventually declassified, other articles were only published as part of collected papers [32]. Classical detonation and shock physics is a phenomenological theory that does not invoke any molecular models or fit parameters. This understanding is almost trivial in the field of acoustics, however, it remains a major hurdle in the wider acceptance of these ideas in molecular biology. Any new physical theory is supposed to be judged against the observation associated with a phenomenon, not the previous interpretations of the observations based on a popular theory. In fact, it provides a new perspective to look at old observations and find new more general interpretations and a deeper understanding.

To further explore the nonlinear nature of action potential from shock physics perspective, experiments on the local nature of excitation, propagation characteristics of the sub and suprathreshold pulses close to excitation – where the nonlinearity is still developing – will be crucial. For example, measurements of the action potential as a function of distance near the electrodes have shown steepening of the pulse similar to a shockwave [77]. The sharp all-or-none behavior exists *only* at a distance from the electrode as the wavefront gets continuously steeper. Similarly, it has also been reported that the velocity and pulse shape depends on the stimulus near the excitation electrode, the velocity usually increases with the excitation strength for a squid axon before setting to a lower value at 1-5 resting length constants [78].

Irrespective of all the typical indications of a shock wave that are associated with a nerve impulse, Eq. (1) and (2) effectively state the conservation of mass, momentum, energy and the equation of state, which are all indisputable and are applicable to all propagative phenomenon. For example, in analogy to Hodgkin and Huxley model, a flame front can also propagate while depleting a chemical reservoir at a subsonic speed, a process known as deflagration in contrast to detonation [35]. Eq. (1) and (2) are still applicable however the velocity, in this case, is further limited by the heat capacity and thermal conductivity of the material, which still don't appear in the Hodgkin and Huxley model. Therefore, while many propagative phenomena in biology could be akin to the deflagration process (spreading depression being a potential candidate), their foundation in Hodgkin and Huxley model and not in eq. (1) and (2) would remain unsatisfactory.

**Conclusion**

It was shown that shock waves observed in lipid monolayer provide evidence of detonation waves in such systems. This opens the doors to investigate a range of acoustic phenomenon in such systems (shock waves, solitons, detonation, deflagration (flame propagation))[28,35], which may or may not have a biological purpose and function. This will require measurements of relevant membrane properties under dynamic conditions, as indicated by the understanding developed here for simple systems like lipid monolayers. The qualitative theory presented here outlines the parameters that need to be measured for a more quantitative understanding, which will be followed by the subsequent development of a quantitative theory.

In conclusion, just like in plants and metals, the phenomenon of response in lipid interfaces is similar to that of nerve impulses in animals. Nonlinearity in the state diagrams of the interface play a critical role in this response and lipids allow us to control these nonlinearities in a convenient manner via phase transitions. Here we pursued the consequences of phase transitions and shock physics and we arrived at the possibility of detonation, i.e. reinforcement of the propagating wave by the latent heat of transition. This was found to be in agreement with previous observations in the lipid interface. The insights from the role of phase transitions in shock waves in lipid interfaces were applied to the pressure and temperature dependence of nerve impulses to derive a possible role for the state diagrams in the phenomenon. Evidence from shock physics as well as the phenomenon of excitation in living as well as nonliving system compel us to reject any theory that gives a special status to living systems as far as excitability is concerned. While being pragmatic about the role of proteins and the associated phenomenon, the ionic hypothesis has to be rejected on similar grounds as envisioned in the concluding remarks by J.C Bose in 1902, "*Nowhere in the entire range of these response phenomena -- inclusive as that is of metals, plants and animals--do we detect any breach of continuity. In the study of processes apparently so complex as those of irritability, we must, of course, expect to be confronted with many difficulties. But if these are to be overcome, they, like others, must be faced and their investigation patiently pursued, without the postulation of special forces whose convenient property it is to meet all emergencies in virtue of their vagueness.*" [44]

**Acknowledgement**

I would like to thank Dr. Konrad Kauffman (Gottingen) for several sessions and discussions on the thermodynamic origin of nerve pulse propagation and its theoretical explanation which laid the foundation for this work. Special thanks to Dr. Daniel Backman for a critical reading of the manuscript. I would also like to thank Prof. Matthias Schneider, Dr. Christian Fillafer and Prof. Robin Cleveland for helpful discussions.

Detonation waves in lipid monolayers                                                                Shamit Shrivastava
46. Shrivastava S, Kang K, Schneider MF. 2015 Solitary Shock Waves and Adiabatic Phase Transitions Lipid Interfaces and Nerves. *Phys. Rev. E* **91**, 12715.

47. Thompson PA, Carofano GC, Kim Y-G. 1986 Shock waves and phase changes in a large-heat-capacity fluid emerging from a tube. *J. Fluid Mech.* **166**, 57–92. (doi:10.1017/S0022112086000046)

48. Tasaki I. 1959 Demonstration of two stable states of the nerve membrane in potassium-rich media. *J. Physiol.* **148**, 306–331. (doi:10.1113/jphysiol.1959.sp006290)

49. Inoue I, Kobatake Y, Tasaki I. 1973 Excitability, instability and phase transitions in squid axon membrane under internal perfusion with dilute salt solutions. *BBA - Biomembr.* **307**, 471–477. (doi:10.1016/0005-2736(73)90294-0)

50. Ueda T, Muratsugu M, Inoue I, Kobatake Y. 1974 Structural changes of excitable membrane formed on the surface of protoplasmic drops isolated from Nitella. *J. Membr. Biol.* **18**, 177–86.

51. Guttman R. 1966 Temperature characteristics of excitation in space-clamped squid axons. *J. Gen. Physiol.* **49**, 1007–1018.

52. Kendig J. 1978 Pressure, temperature, and repetitive impulse generation in crustacean axons. *J. Appl. Physiol.* **45**, 742–746.

53. Fillafer C, Schneider MF. 2013 On the Temperature Behavior of Pulse Propagation and Relaxation in Worms, Nerves and Gels. *PLoS One* **8**, e66773. (doi:10.1371/journal.pone.0066773)

54. Tasaki I, Carbone E, Sisco K, Singer I. 1973 Spectral analyses of extrinsic fluorescence of the nerve membrane labeled with aminonaphthalene derivatives. *Biochim. Biophys. Acta (BBA …* **323**, 220–233.

55. Luzzati V, Mateu L, Marquez G, Borgo M. 1999 Structural and electrophysiological effects of local anesthetics and of low temperature on myelinated nerves: implication of the lipid chains in nerve excitability. *J. Mol. Biol.* **286**, 1389–402. (doi:10.1006/jmbi.1998.2587)

56. Georgescauld D, Desmazes J, Duclohier H. 1979 Temperature dependence of the fluorescence of pyrene labeled crab nerve membranes. *Mol. Cell. Biochem.* **27**, 147–153.

57. Georgescauld D, Duclohier H. 1978 Transient fluorescence signals from pyrene labeled pike nerves during action potential possible implications for membrane fluidity changes. *Biochem.*
26